# TWO-ION DUSTY PLASMA WAVES AND LANDAU DAMPING

## B. ATAMANIUK and K.ŻUCHOWSKI (WARSZAWA)

*Institute of Fundamental Technological Research, PAS, Świętokrzyska 21, 00-049, Warsaw, Poland.*

The paper analyses the properties of dusty plasmas in the extreme conditions when the free electrons are absent. The nonlinear Korteveg de Vries equation with a nonlocal (integral) term in a small parameter approximation is derived. The conditions are determined when the integral term is essential hence the Landau damping of two-ion-dusty plasma waves is substantial.

## 1. Introduction.

Plasmas with dust grains are of interest both for the cosmic space as well as for the laboratory plasmas. Examples include cometary environments, planetary rings, the interstellar medium, and the earth's magnetosphere [1]. Dust has been found to be a determinant component of rf plasmas used in the microelectronic processing industry, and it may also be present in the limiter regions of fusion plasmas due to sputtering of the carbon by energetic particles. It is interesting to note the recent flurry of activity in the dusty plasma research. It has been driven largely by discoveries of the role of dust in quite different settings: the ring of Saturn [1] and the plasma processing device [2]. Dusty plasmas contain, beside positive ions and electrons, big particles usually negatively charged. They are conglomerations of the ions, electrons and neutral particles. These big particles, to be called grains, have atomic numbers $Z_d$ in the range of $10^4 - 10^6$ and their mass $m_d$ can be equal to $10^6$ of proton mass or more. It is assumed, for simplicity, that all grains have equal masses and charges, which are practically constant, and the collisions of all particles are neglected in the considered interval of time. We note that for dusty plasmas, ratio of electrical charges of grains to it masses is usually much smaller than in the case of negative ions [3]. The ratio of charge to mass for a given component of the plasma determines its dynamics [4].

In the considered dusty plasmas the size of grains is small compared with average distance between grains. Because dynamics of the dusty plasma components, electrons, positive ions and dust grains, is quite different in the time and length scales considered here, then the equations for the aforementioned components of the dusty plasma may be different. We use fluid and Vlasov descriptions for appropriate components of the dusty plasma.

## 2. Dispersion relation for acoustic waves.

In plasma, where the dust is absent, ion-acoustic waves occur if the electron temperature is much greater than the ion temperature. For the waves, in a longitudinal approximation, the phase velocity is, see [4]:

$$\left(k_B T_e / m_i\right)^{1/2},$$

where: $k_B, T_e$ and $m_i$ denote Boltzmann's constant, electron temperature and ion mass, respectively.

For the dusty plasma, if the phase velocity of the wave is much greater than the thermal velocity of ions and dust but much smaller than the thermal velocity of electrons, we have, in a longitudinal approximation, the following relation for the phase velocity [5]:

$$\frac{\omega}{k} = \left[\frac{n_{i0}}{n_{e0}}\left(\frac{k_B T_i}{m_i}\right)\right]^{1/2},$$

where $\omega, k, n_{i0}$ and $n_{e0}$ denote the wave frequency , the wave number, and the unperturbed ion and electron densities, respectively.

The dusty-plasmas admit the waves as the temperature of ions $T_i$ and the temperature of electrons $T_e$ is nearly equal, $T_i \approx T_e$. The presence of dust grains can have a profound influence on low frequency waves leading to the phase velocity of the waves much smaller than the thermal velocity of electrons and ions, but much greater than the thermal velocity of the dust. For that dust-acoustic waves (DAW), in a longitudinal approximation, we have the following phase velocity, see [6]:

$$\omega / k = \beta v_d,$$

where $\beta^2 = Z_d (\delta - 1) / (1 + \gamma \delta)$, $v_d^2 = k_B T_e / m_d$, $\delta = n_{i0} / n_{e0}$ and $\gamma = T_e / T_i$.

Fluid approach was used for the above described dust-plasma waves [5] and DAW [6]. The same waves were obtained from the Vlasov-Poisson system of equations, [7]. As usually, the global charge neutrality was assumed: $n_{i0} = Z_d n_{d0} + n_{e0}$, where $n_{d0}$ denotes the unperturbed dust density. We treat the special case when free electrons are absent from the considered dusty plasmas and then the condition of global charge neutrality takes the form: $n_{i0} = Z_d n_{d0}$. In that case, the phase velocity of the dust acoustic wave, in a longitudinal approximation, takes the form:

$$\omega / k = \left( Z_d T_i / m_d \right)^{1/2}.$$

### 3. Dust-Ion Plasmas.

Recently, attention has been focused on planetary rings in which the heavy, micron-sized dust grains are charged to the high degree voltage. In particular, in the F-ring of Saturn there is an anomalous situation where the number density of free electrons is much smaller than the number density of ions [1]. This could happen because the charged dust grains collect electrons from the background medium. We consider an extreme situation when the free electrons are absent. The cold dust (negatively charged) is described by fluid equations (3.1), (3.2). Two species of ions: cold and hot singly ionized of equal masses $m_i$ are described by the kinetic Vlasov equation (3.4), whereas the electric potential $\phi$ is described by the Poisson equation (3.3) as follows:

$$(3.1) \qquad \frac{\partial n_d}{\partial t} + \frac{\partial}{\partial x}\left( n_d u_d \right) = 0,$$

$$(3.2) \qquad \frac{m_d}{Z_d}\left( \frac{\partial u_d}{\partial t} + u_d\, \frac{\partial u_d}{\partial x} \right) = e\, \frac{\partial \phi}{\partial x},$$

$$(3.3) \qquad \partial^2 \phi / \partial x^2 = \frac{e}{\varepsilon_0}(Z_d n_d - n_i)\,,$$

$$(3.4) \qquad \frac{\partial f_{ij}}{\partial t} + v\frac{\partial f_{ij}}{\partial x} - \frac{e}{m_i}\frac{\partial \phi}{\partial x}\frac{\partial f_{ij}}{\partial v} = 0\,,$$

where the index $j$ denote $c$ or $h$ (cold or hot ions). The total ion density is given by:

$$(3.5) \qquad n_i = \int_{-\infty}^{\infty} (f_{ic} + f_{ih}) dv.$$

Moreover, in (3.1)-(3.5) $n_d, u_d, \varepsilon_0, e, f_{ic}$ and $f_{ih}$ denote the dust number density, the dust fluid velocity, the free space permittivity, the positive elementary charge, the velocity distribution function for the cold and hot ions, respectively. Next, we normalize all quantities to be dimensionless: dust density to equilibrium value $n_{d0} = n_{i0} / Z_d$, potential $\phi$ to $k_B T_{ic} / e$ (where $T_{ic}$ denotes temperature of cold ions), length to the characteristic scale wave length $L$, dust fluid velocity to $a_0 = (k_B T_c Z_d / m_d)^{1/2}$, ion velocity to the ion thermal velocity $a_i = (k_B T_{ic} / m_i)^{1/2}$, time to $L / a_0$, and ion distribution function to $n_{i0}$. Next, we introduce transformation of the dependent variables:

$Z_d n_d = n$, $\phi = -\phi'$ and $m_d / Z_d = m$.

Introducing this dimensionless variable, we have

$$(3.1a) \qquad \frac{\partial n}{\partial t} + \frac{\partial}{\partial x}(n u_d) = 0,$$

$$(3.2b) \qquad \frac{\partial u_d}{\partial t} + u_d \frac{\partial u_d}{\partial x} = -\frac{\partial \phi'}{\partial x},$$

$$(3.3c) \qquad \frac{\lambda_D^2}{L^2} \frac{\partial^2 \phi'}{\partial x^2} = n_i - n,$$

$$(3.4d) \qquad \left(\frac{m_i}{m}\right)^{1/2} \frac{\partial f_{ij}}{\partial t} + v \frac{\partial f_{ij}}{\partial x} + \frac{\partial \phi'}{\partial x} \frac{\partial f_{ij}}{\partial v} = 0,$$

where $\lambda_D = \left(\varepsilon_0 k_B T_{ic} / n_{d0} Z_d e^2\right)^{1/2}$ is the Debye length, which is defined by use of the cold ion temperature.

Now, following [8], we transform the coordinate into a moving wave frame and introduce the stretching in time:

(3.6)
$$\xi = x - \frac{a_{s0}}{a_0}t, \quad \tau = \varepsilon t,$$

where $a_{s0}$ is the dust acoustic speed for the two-ion-temperature plasma:

(3.7)
$$a_{s0} = \left[ \frac{k_B T_{ic} Z_d e^2}{m_d} \frac{1 + \frac{n_{i0}^h}{n_{i0}^c}}{1 + \frac{T_{ic}}{T_{ih}} \frac{n_{i0}^h}{n_{i0}^c}} \right]^{1/2},$$

(3.8)    $T_{ih}$ denotes the temperature of hot ions. The physical quantities are expanded in power series with respect to $\varepsilon$ in the relation to the equilibrium state, as follows:

(3.9)
$$n = 1 + \varepsilon n^{(1)} + \varepsilon^2 n^{(2)} + \varepsilon^3 n^{(3)} + \cdots,$$
$$u_d = + \varepsilon u_d^{(1)} + \varepsilon^2 u_d^{(2)} + \varepsilon^3 u_d^{(3)} + \cdots,$$
$$\phi' = \varepsilon \phi'^{(1)} + \varepsilon^2 \phi'^{(2)} + \varepsilon^3 \phi'^{(3)} + \cdots,$$
$$n_{ic} = \frac{n_{i0}^c}{n_{d0} Z_d} + \varepsilon n_{ic}^{(1)} + \varepsilon^2 n_{ic}^{(2)} + \varepsilon^3 n_{ic}^{(3)} + \cdots,$$
$$n_{ih} = \frac{n_{io}^h}{n_{d0} Z_d} + \varepsilon n_{ih}^{(1)} + \varepsilon^2 n_{ih}^{(2)} + \varepsilon^3 n_{ih}^{(3)} + \cdots,$$
$$f_{ic} = f_{ic}^{(0)} + \varepsilon f_{ic}^{(1)} + \varepsilon^2 f_{ic}^{(2)} + \varepsilon^3 f_{ic}^{(3)} + \cdots,$$
$$f_{ih} = f_{ih}^{(0)} + \varepsilon f_{ih}^{(1)} + \varepsilon^2 f_{ih}^{(2)} + \varepsilon^3 f_{ih}^{(3)} + \cdots,$$

where $f_{ic}^{(0)}$ and $f_{ih}^{(0)}$ are assumed to be the following Maxwellian velocity distributions:

(3.9)
$$f_{ic}^{(0)} = \frac{n_{i0}^c}{n_{d0} Z_d} \frac{1}{\sqrt{2\pi}} \exp\left( -\frac{v^2}{2} \right),$$

(3.10)
$$f_{ih}^{(0)} = \frac{n_{i0}^h}{n_{d0} Z_d} \sqrt{\frac{T_{ic}}{2\pi T_{ih}}} \exp\left( -\frac{1}{2} \frac{T_{ic}}{T_{ih}} v^2 \right),$$

where $n_{i0}^c + n_{i0}^h = n_{i0}$. We now consider amplitude waves in the two-ion-temperature dusty plasma with no electrons. We expect that the Landau damping of the waves may by important in such a case. Following [9] and [8], we assume

(3.11)
$$(\frac{m_i}{m})^{1/2} = Z_d^{1/2}\left(\frac{m_i}{m_d}\right)^{1/2} = \alpha_1\varepsilon,$$

and

(3.12)
$$\frac{\lambda_D^2}{L^2} = \alpha_2\varepsilon,$$

where $(m_i/m)^{1/2} = Z_d^{1/2}(m_i/m_d)^{1/2}$ and $\lambda_D^2/L^2$ are the measures of the Landau damping strength of ions and dispersion effects, respectively.
Substituting Eqs. (3.6), (3.8) and (3.11) into Eqs. (3.1a)-(3.4d) and (3.5) (after normalization), then equating terms at the powers of $\varepsilon$ to zero, we obtain the infinite set of equations. . Performing the calculations in a similar way as in [8] and [9], we obtain the following relations for the first approximation:

(3.13)
$$u_d^{(1)} = \frac{a_0}{a_{s0}}\phi'^{(1)},$$

(3.14)
$$n^{(1)} = \frac{a_0^2}{a_{s0}^2}\phi'^{(1)},$$

(3.15)
$$n^{(1)} - n_i^{(1)} = 0,$$

(3.16)
$$f_{ic}^{(1)} = f_{ic}^{(0)}\phi'^{(1)},$$

(3.17)
$$f_{ic}^{(1)} = \frac{T_{ic}}{T_{ih}}f_{ih}^{(0)}\phi'^{(1)}.$$

The second order quantities at $\varepsilon$ provide the equation:

(3.18)
$$\frac{\partial n^{(1)}}{\partial \tau} + an^{(1)}\frac{\partial n^{(1)}}{\partial \xi} + \alpha_1 bP\int\frac{\partial n^{(1)}}{\partial \xi'}\frac{d\xi'}{\xi - \xi'} + \alpha_2 c\frac{\partial^3 n^{(1)}}{\partial \xi^3} = 0,$$

where

$$(3.19) \qquad a = \frac{1}{2} \frac{a_{s0}}{a_0} \left[ 3 - \frac{n_{i0}\left(\dfrac{n_{i0}^c}{T_{ic}^2} + \dfrac{n_{i0}^h}{T_{ih}^2}\right)}{\left(\dfrac{n_{i0}^c}{T_{ic}} + \dfrac{n_{i0}^h}{T_{ih}}\right)^2} \right],$$

$$(3.19a) \qquad b = \frac{1}{2} \frac{a_{s0}^4}{a_0^4} \frac{n_{i0}^c}{n_{i0}} \left[ 1 + \frac{n_{i0}^h}{n_{i0}^c}\left(\frac{T_{ic}}{T_{ih}}\right)^{3/2} \right],$$

$$(3.20) \qquad c = \frac{1}{2} \frac{a_{s0}^3}{a_0^3}.$$

In (3.18), P denotes the Cauchy principal value, and to derive the equation we made use of the global charge neutrality condition: $Z_d n_{d0} = n_{i0}^c + n_{i0}^h$.

The equation (3.18) is the Korteweg-de Vries type equation with nonlocal integral term and it was previously derived in [9] and [8] for the ion acoustic waves in one-electron-temperature and two-electron-temperature plasmas, respectively. Now, we investigate the range of the validity of (3.18) equation.

If $\alpha_1 \ll 1$ and $\alpha_2 \sim O(1)$, then

$$1 \gg \Delta n / Z_d n_{d0} \approx \lambda_D^2 / L^2 \gg \left(\frac{Z_d m_i}{m_d}\right)^{1/2}$$

and we can neglect the nonlocal term due to the Landau damping effect on the ions, and then the Eq. (3.18) reduces to the ordinary Korteweg-de Vries equation.

However, for such a small wave amplitude as

$$\Delta n / Z_d n_{d0} \approx \lambda_D^2 / L^2 \approx \left(Z_d m_i / m_d\right)^{1/2},$$

the Landau damping effect becomes important since the nonlocal term can not be neglected.

We introduce the parameter D given by

$$(3.21) \qquad D = 3 - \frac{\left(n_{i0}^c + n_{i0}^h\right)\left(\dfrac{n_{i0}^c}{T_{ic}^2} + \dfrac{n_{i0}^h}{T_{ih}^2}\right)}{\left(\dfrac{n_{i0}^c}{T_{ic}} + \dfrac{n_{i0}^h}{T_{ih}}\right)^2},$$

which has an essential influence on the form of nonlinear term in Eq. (3.18) because $a = \dfrac{1}{2}\dfrac{a_{s0}}{a_0}D$.

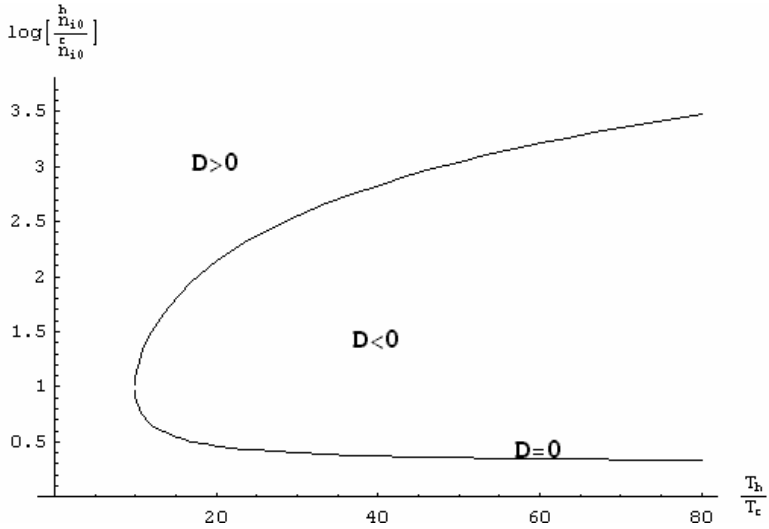

**Fig. 1**. The D-plane of normal (D>0) and anomalous (D<0) regions.

In the Fig. 1, we show the D- plane in the coordinates $T_{ih}/T_{ic}$ and $\log\left[n_{i0}^h/n_{i0}^c\right]$. The plane is divided into two regions by the curve D=0, that is the normal region for D>0 and the anomalous region for D<0. For D=0, the nonlinear term in Eq. (3.18) vanishes. We note that in the case D=O(1), the transformations (3.6), (3.11), and (3.12) are valid. But in the case of D=0($\varepsilon$), the referred transformations are to be changed by the substitution $\varepsilon \to \varepsilon^2$. Consequently, the Eqs.(3.1a)-(3.4d) and (3.5) have to be expanded to the higher order accuracy with respect to $\varepsilon$, which is to include $\varepsilon^3$. Then the first order density $n^{(1)}$, which is simply related to the first order dust density $n^{(1)} = Z_d n_d^{(1)}$, is to be described by the equation in which appears the higher order nonlinear term in comparison to the eq. (3.18). In the transformations, mentioned above the relationship between the first order quantities in $\varepsilon$ are of the same form as in (3.14)-(3.17). We also note that the first order perturbations of dust density and electric potential are of opposite signs.

Another important conclusion is that the Landau damping can occur for ions as

$$\lambda_D^2 / L^2 \approx \left(\frac{Z_d m_i}{m_d}\right)^{1/2} \text{ or even } \left(\frac{Z_d m_i}{m_d}\right)^{1/2} >> \lambda_D^2 / L^2,$$

and then the Landau damping is dominant.

## Acknowledgments

The authors are grateful to Doc. A. J. Turski for critical reading of the manuscript.

The paper was partly supported by the State Committee for Scientific Research KBN through the grant No 2P03C01210.

## References.

INSTITUTE OF FUNDAMENTAL TECHNOLOGICAL RESEARCH, PAS,
ŚWIĘTOKRZYSKA 21, 00-049, WARSAW, POLAND.
E-mail: batama@ippt.gov.pl
        kzucho@ippt.gov.pl